\documentclass{aa}
\usepackage{graphicx}
\usepackage{txfonts}                   
\usepackage{placeins}          
\usepackage{makecell}
\makeatletter

\newcommand{\Rmnum}[1]{\expandafter\@slowromancap\romannumeral #1@}
\newcommand{\kms}{km\,s$^{-1}$}

\makeatother                                
\usepackage[colorlinks=true,allcolors=blue]{hyperref}

\begin{document}

   \title{Statistics of transition-region loop brightenings and their heating implication}

   \author{Xiuhui Zuo\inst{1}
        \and Zhenghua Huang\inst{2}\fnmsep\inst{3}\fnmsep\thanks{huangzh@nju.edu.cn}
        \and Maria S. Madjarska\inst{4}\fnmsep\inst{5}\fnmsep\inst{6}
        \and Hui Fu\inst{1}
        \and Hengyuan Wei\inst{2}\fnmsep\inst{3}
        \and Xinzheng Shi\inst{1}
        \and Lidong Xia\inst{1}
        }

   \institute{Shandong Key Laboratory of Space Environment and Exploration Technology, Institute of Space Sciences, School of Space Science and Technology, Shandong University, Shandong, China
   \and Institute of Science and Technology for Deep Space Exploration, Suzhou Campus, Nanjing University, Suzhou 215163, China
   \and State Key Laboratory of Lunar and Planetary Sciences, Macau University of Science and Technology, Macau, China
   \and Max Planck Institute for Solar System Research, Justus-von-Liebig-Weg 3, 37077, G\"ottingen, Germany
   \and Space Research and Technology Institute, Bulgarian Academy of Sciences, Acad. G. Bonchev Str., Bl. 1, 1113, Sofia, Bulgaria
   \and Korea Astronomy and Space Science Institute, 34055, Daejeon, Republic of Korea}
   
   \authorrunning{Zuo et al.}
   \date{Received ....; accepted ....}

  \abstract
   {Transition-region loops are a type of critical magnetic structure in the solar atmosphere, yet their physical properties and evolutionary characteristics remain statistically poorly constrained.}
  {We aim to statistically characterize the physical properties of propagating brightening events in transition-region loops and to explore the underlying heating mechanism responsible for these brightenings.} 
   {Using coordinated observations from the Extreme Ultraviolet Imager onboard the Solar Orbiter and the Atmospheric Imaging Assembly (AIA) onboard the Solar Dynamics Observatory, we analyze 42 propagating brightening events in loops that are unambiguously detected in both instrument data.
   Each of these events evolve simultaneously in the AIA 94, 131, 171, 193, 211, 304, and 335~\AA\ passband images, suggesting that they are in the transition-region or low-coronal temperature range. }
   {Our analyses show that these brightenings are impulsive, with an average brightening time of
   	$118.4 \pm 12.0~\mathrm{s}$ and a mean intensity decreasing time of $159.4 \pm 16.6~\mathrm{s}$. The propagating brightenings are predominantly subsonic, with velocities in the range of 0--90~\kms\ and an average of $51.3 \pm 5.6$~\kms.
   	The lengths of brightenings range from 3 to 11~Mm, with an average and standard deviation of $6.3 \pm 0.4~\mathrm{Mm}$, which are closely related to the propagation velocity and the lifetime.
	The initial brightening sites are predominantly located near the footpoints of these loops, and the number of brightening events decreases systematically with increasing of loop height. }
   {Our results are consistent with an energizing mechanism regulated by enthalpy flows and radiative cooling. 
   	Dynamic magnetic evolution, including migration and cancellation, are taking place at the footpoints of these loops, and thus we suggest that magnetic cancellation, driven by reconnection and/or braiding that generate these brightenings, is a possible mechanism of plasma heating in a transition-region loop.
   	Based on the measurements of these brightenings, we also propose a diagnostic method for the temperature or density of transition-region loops.}


   \maketitle
   \nolinenumbers

\section{Introduction}
In the one-dimensional static model, the solar atmosphere is commonly described as a stratified system consisting of the photosphere, chromosphere, transition region, and corona. The photosphere is the visible solar surface; the chromosphere is the overlying layer where temperature increases gradually; and the corona is the hot outer atmosphere characterized by temperatures of the order of 1~MK and above. The transition region is the interface between the chromosphere and the corona, characterized by a rapid temperature rise over a relatively short height range. It is normally defined in the temperature regime from roughly $2\times10^4$\,K to $8\times10^5$\,K. One of the major unresolved problems in solar physics is why the temperature rises sharply from the lower atmosphere to the corona, namely the coronal heating problem\,\citep{edlen1940attempt,10.1093/mnras/107.2.211,1993SoPh..148...43Z,2006SoPh..234...41K,2015RSPTA.37340256K}. 
Loops are arcade-like feature in the solar atmosphere especially the corona, which are the results of the coupling between the solar magnetic field and the hot solar plasma.
They are one of the fundamental structures of the solar atmosphere, and thus they are an ideal object to study and resolve the coronal heating problem\,\citep{2003A&A...406.1089D,2014LRSP...11....4R,2015ApJ...810...46H,2015SSRv..188..211F}. Based on the temperature, loops are usually classified as cool ($<$1~MK), warm (1--2~MK), and hot loops ($>$2~MK)\,\citep{2014LRSP...11....4R}. In terms of heating timescales, heating is commonly divided into impulsive heating and steady heating. Independent heating events with heating times shorter than the cooling times are regarded as impulsive heating\,\citep{2004ApJ...605..911C,2011ApJ...734...90W}, whereas heating with timescales much longer than the cooling times, or the case where numerous impulsive events occur in rapid succession\,\citep{2024NatAs...8..706L}, is treated as steady heating\,\citep{2015A&A...583A.109L}. The density and magnetic field strength of coronal loops usually decrease with height along the loop\,\citep{2001SoPh..203...71G,2017ApJ...842...38X}. The electron density in loops at coronal temperatures is generally of the order $10^{9}$~cm$^{-3}$ or lower\,\citep{2002ApJ...580..566R,2008ApJ...680.1477A,2018ApJ...869..175H}.

\par
The heating distribution along loops, i.e., the location of energy deposition, provides a key constraint for establishing the heating mechanism(s) at work. Numerical modeling shows that, for different choices of heating duration, power, and frequency, footpoint heating, approximately uniform heating, or apex heating, can reproduce various observed properties\,\citep{2000ApJ...539.1002P,2000ApJ...541.1059A,2001ApJ...550.1036A,2001ApJ...559L.171A,2002ApJ...580..566R,2011ApJ...734...90W}. High-resolution observations further reveal that mixed polarity magnetic field and their activity, e.g., cancellation, near loop footpoints likely play a key role in heating the plasma to transition-region and coronal temperatures, and that the actual heating often exhibits multi-stranded, intermittent, and highly non-uniform spatial characteristics\,\citep{2017ApJS..229....4C,2018A&A...615L...9C,2019ApJ...887..221H}.

\par
In addition to long and high coronal loops, active regions are also populated by numerous short, low-lying loops, many of which are in the transition-region to low-coronal temperature range.
Here we refer these loops with transition-region temperatures as transition-region loops. 
Transition-region loops are normally have a compact morphology, rapid evolution, and multi-thermal EUV response,
and these are consistent with relatively cool plasma rather than hot coronal loops\,\citep{2015ApJ...810...46H}. 
The density, temperature, and dynamical properties of transition-region loops differ significantly from those of coronal loops. Because of the scarcity of suitable observations, transition-region loops have not been well studied. 
Since the advent of Interface Region Imaging Spectrograph\,\cite[IRIS,][]{2014SoPh..289.2733D}, a number of studies have investigated the physical parameters of transition-region loops. IRIS Si~\textsc{iv} 1400~\AA\ slit-jaw images and spectra directly resolve numerous, highly dynamic, low-lying transition-region loops\,\citep{2014Sci...346E.315H,2018A&A...611L...6P}. 
These loops typically have lengths of only a few megameters, heights of about 1--4~Mm, lifetimes of several minutes, and evolution timescales of tens to a few hundreds of seconds\,\citep{2014Sci...346E.315H,2018A&A...611L...6P,2025A&A...699A..61D}. Many studies showed that transition-region loops have peak temperatures generally below 0.5~MK, while their densities can reach $10^{11}$~cm$^{-3}$ or even larger, significantly higher than those of coronal loops \citep{1998SoPh..182...73K,2015ApJ...810...46H,2024ApJ...966L...6L}. \citet{2014Sci...346E.315H} pointed out that such structures can be driven by intermittent heating associated with magnetic braiding.

\par
Transition-region loops are highly dynamic\,\citep{1998SoPh..182...73K}, and are possibly accompanied by a variety of small-scale energy release processes, and their dynamics plays a crucial role in the transport of mass and energy between the chromosphere and corona\,\citep{2017MNRAS.464.1753H,2019STP.....5b..58H}. Early spectroscopic studies already revealed flows, oscillations, and explosive events in the transition region\,\citep{1989SoPh..123...41D,1998SoPh..182...73K,1999ApJ...511L.121B}. Using IRIS data, \citet{2015ApJ...810...46H} found siphon flows in some cool transition-region loops, with Doppler velocities varying continuously along the loop from about 10~\kms\ at one footpoint to about 20~\kms\ at the other. Many studies based on IRIS data showed that many transition-region loops are closely associated with small-scale activities such as explosive events (EEs) and ultraviolet (UV) bursts\,\citep[e.g.,][and references therein]{2015ApJ...810...46H,2019STP.....5b..58H,2021Symm...13.1390H}. 
Other studies indicate that the dynamics of transition-region loops can lead to small-scale brightenings in moss regions\,\citep{2024ApJ...977...25R}, where propagating bright fronts along the loops are clearly visible, with apparent velocity of the order tens of \kms\,\citep{2021Symm...13.1390H}. Simulations of transition-region loops suggest that reproducing the observed IRIS signatures requires highly intermittent heating in time (multiple, high-frequency events), multi-scale heating, and strongly non-equilibrium ionization\,\citep{2024ApJ...971...59B}. MHD simulations show that random motions of magnetic footpoints in a bipolar system can generate numerous small-scale loops heated to transition-region temperature, whose lifetimes and kinematics are comparable to the observation events\,\citep{2023A&A...672A..47S}.

\par
Current studies of activities in transition-region loops, however, focus on individual events, such as explosive events (EEs) and ultraviolet (UV) bursts. Systematic constraints on the statistical properties of brightenings in transition-region loops in active regions, such as the length of the brightening segment ($L_b$), their propagation velocity along the loops, and heating and cooling times, remain scarce. Our statistical analysis is based primarily on the high spatial and temporal resolution data from the Extreme Ultraviolet Imager\,\citep[EUI,][]{2020A&A...642A...8R} in the 174~\AA\ passband. We investigate 42 brightening events (internal fine-scale activity) in transition-region loops. We compile the relevant physical parameters describing their dynamics, including brightening durations, $L_b$, and the propagation velocity of brightening fronts, and based on the magnetic field evolution in this region, we suggest possible physical mechanisms responsible for these brightening events. These results are of significant value for current coronal heating studies, especially for understanding the heating mechanisms of transition-region loops, and provide key constraints for future related studies. The remainder of this paper is organized as follows. Section~\ref{sec:obs} presents the observations. Section~\ref{sec:res} describes the data analysis and results. Section~\ref{sec:dis} provides the discussion. Section~\ref{sec:con} summarizes our conclusions.

\begin{figure*}[!t]
	\centering
	\includegraphics[width=\textwidth]{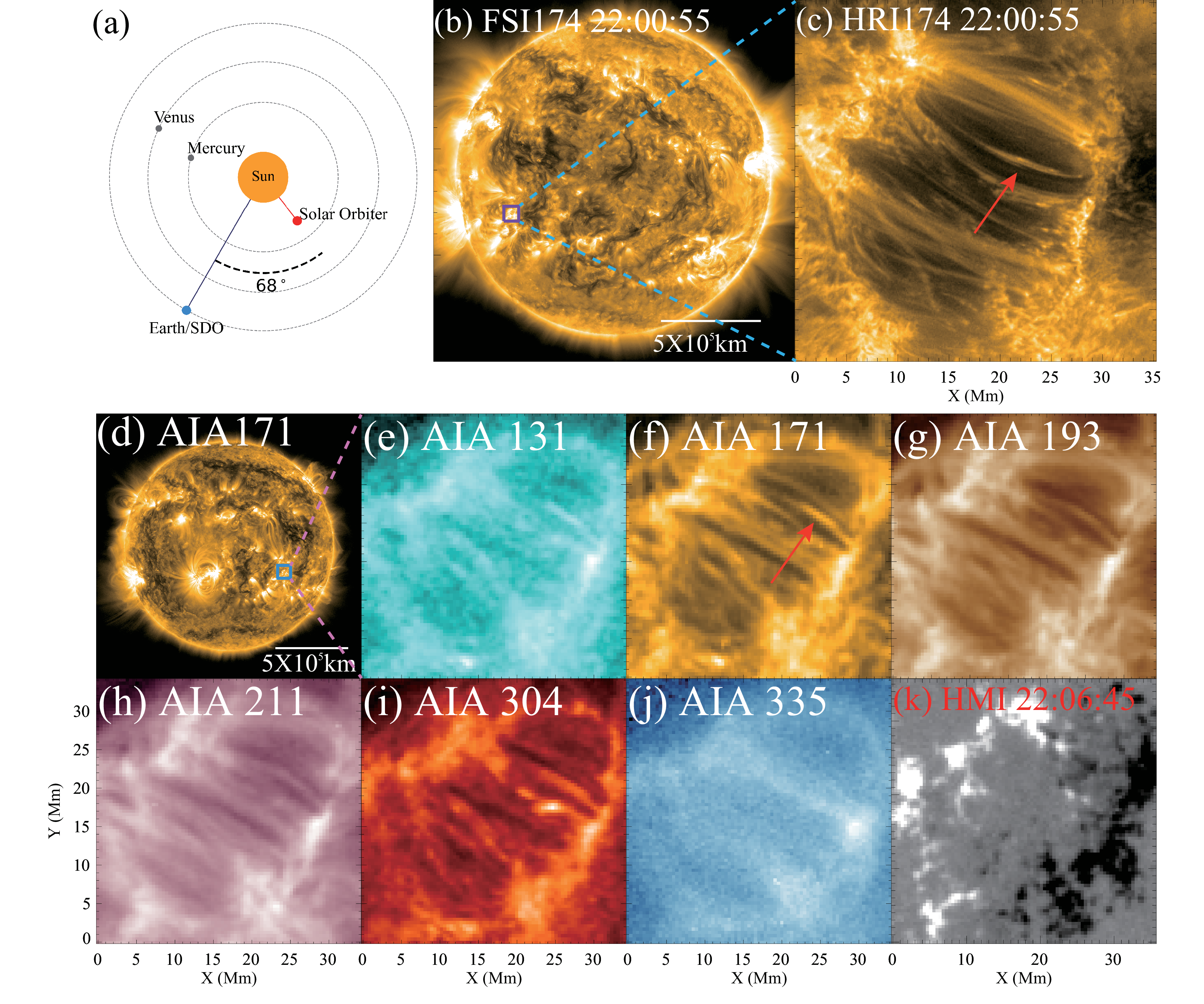}
	\caption{Overview of the study. Panel (a) shows the observation positions of SDO and SO. 
		Panel (b) presents the full disk of the sun in EUI 174~\AA\ passband, of which the purple box indicates the ROI location in the HRI field of view, situated near the equatorial region on the left side of the solar surface. Panel (c) is the ROI in a high-resolution image of HRI$_{\mathrm{EUV}}$ passband. Panel 
		(d) shows the full disk of the Sun in AIA~171~\AA\ passband, of which the blue box marks the ROI position in the AIA field of view, located near the equatorial region on the right side of the solar surface. Panels 
		(e)–(j) display the ROI images of AIA passbands, where multiple distinct loops connect the magnetic polarity regions in the upper left and lower right. 
		Panel (k) shows the HMI magnetic field image of the ROI scaled from $-$300~G (black) to 300~G (white), with the primary positive polarity region in the upper left field of view, the main negative polarity region in the lower right, and numerous small-scale polarity points distributed in the central region. }
	
	\label{fig1}
\end{figure*}

\section{Observations} \label{sec:obs}
In this study, we utilize observations from the Atmospheric Imaging Assembly\,\citep[AIA,][]{2012SoPh..275...17L}  and the Helioseismic and Magnetic Imager \citep[HMI,][]{2012SoPh..275..207S} onboard Solar Dynamics Observatory\,\citep[SDO,][]{2012SoPh..275....3P}, and EUI/Solar Orbiter\,\citep[SO,][]{2020A&A...642A...1M}. 
The EUI instrument consists of three telescopes: the Full Sun Imager (FSI), which operates at 174 and 304~\AA\ passbands, and two High Resolution Imagers, namely HRI$_{\mathrm{EUV}}$ and HRI$_{\rm Ly\alpha}$.
The FSI 174~\AA\ passband provides full-disk EUV images which is used to locate the region of interest (ROI) on the solar disk (Fig.~\ref{fig1}(b)) and the HRI$_{\mathrm{EUV}}$ provides high-resolution data for detailed loop structure analysis (Fig.~\ref{fig1}(c)). 
The observations analyzed here were obtained from the EUI Data Release 6.0\,\citep{euidatarelease6}, and they are a part of the Solar Orbiter Observing Plan (SOOP) named ``R\_BOTH\_HRES\_HCAD\_Nanoflares''.
The data were acquired on 2023 April 11 and focused on the peripheral portion of active region AR~13269. During these observations, SO was located at a heliocentric distance of 0.29~AU. The SO--SDO observing times differ by 360~s, reflecting the light-travel-time offset. The longitudinal separation angle between the two lines of sight is 68$^{\circ}$ (Fig.~\ref{fig1}(a)), and the solar latitude of SO was 2.26$^{\circ}$. In the EUI field of view, the ROI lies slightly to the east of the disk center, whereas in the AIA field of view was slightly to the west of the disk center. This longitudinal separation allows us to view the brightenings from different angles and to reduce the impact of projection effects that may otherwise obscure the structures.
The co-alignment between the SO and SDO data was determined from the relative observing angle of the two instruments and refined through visual comparison of common structures in the ROI, including bright dots, loop footpoints and intersections, etc.\,\citep{2025ApJ...985...17Z}.

\par
The time interval of interest for HRI observations is 21:34--23:34 UT. The HRI data have a spatial resolution of 0.492\arcsec, corresponding to a pixel size of about 110~km, and a temporal cadence of 3~s and 5~s. 
AIA/SDO provides data with a temporal resolution of 12~s and a spatial resolution of 426~km~pixel$^{-1}$. In the HRI$_{\mathrm{EUV}}$ images, the loops are clearly identifiable (Fig.~\ref{fig1}(c)). 
For the events analyzed in this study, the observed loops are identified based on their coherent, continuous, arc-shaped morphology in the HRI$_{\mathrm{EUV}}$ images. Their widths are generally larger than 5 pixels (about 550~km), whereas their projected lengths typically range from several tens to a few hundred pixels. The measured loop widths are significantly larger than the Point Spread Function (PSF) scale that is about 220~km\,\citep{2022A&A...667A.166C}. These characteristic sizes, together with the clear loop-like continuity, support the identification and tracking of the observed structures as loops. This enables us to analyze parameters such as the loop length, $L_b$ (length of brightening segment), propagation velocity, brightening time, and intensity decreasing time. Nevertheless, finer strand-like substructure, if present, may remain unresolved in the current data and will require future observations with higher spatial resolution. 

\par 
The HRI$_{\mathrm{EUV}}$ passband is most sensitive to plasma with temperatures around 1~MK. Because of its relatively broad temperature response, plasma at several $10^{5}$~K also strongly contributes to this passband, providing the physical basis for using these data to study transition-region loops. AIA supplies observations in seven EUV passbands (94, 131, 171, 193, 211, 304, and 335~\AA\ passbands). Although its lower spatial and temporal resolution prevents us from resolving the fine loop morphology in the AIA field of view, these passbands together cover emission lines formed over a wide temperature range from several $\times10^{4}$~K to several $\times10^{7}$~K, which is highly advantageous for our temperature estimation. The HMI data have a spatial resolution of 0.6\arcsec\ per pixel and a temporal cadence of 45~s, and we use the HMI data to investigate the magnetic field evolution of the active region (Fig.~\ref{fig1}(f)), which allows us to analyze the evolution of the magnetic field at the loop footpoint.

\section{Data analysis and results} \label{sec:res}
Figure~\ref{fig1} presents the overview of the ROI. SDO and SO viewed the ROI from different vantage angles (Fig.~\ref{fig1}(a)), and the ROI appears nearly symmetrically located within the fields of view of the two instruments, which helps mitigate projection effects associated with viewing geometry (Figs.~\ref{fig1}(b)\&(d)). As shown in Figs.~\ref{fig1}(c)--(j), owing to the differing perspectives, most loops appear as upward convex arches in the HRI field of view, whereas they appear with an opposite curvature in the AIA view. These loops primarily connect the dominant positive polarity region in the upper left and the dominant negative polarity region in the lower right, while a subset of shorter loops connect within the intermediate zone. To enable uniform quantitative measurements of loop brightening events, we co-aligned the HRI$_{\mathrm{EUV}}$ data with the AIA data in both time and space (Fig.~\ref{fig1}). Brightening events were then identified and marked by visual inspection, and time-slice (T--S) analysis along the loop trajectories in the HRI$_{\mathrm{EUV}}$ field of view was used to extract light curves and to quantify event properties. Following this procedure, we identified 202 brightenings in the HRI data, among which 42 events can be clearly recognized in the AIA field of view and satisfy the requirements for subsequent quantitative analysis (i.e., the loop is unambiguously resolved, the brightening evolution is fully covered in time, and the morphology is not severely affected by complex loop superposition due to projection effect). We therefore perform our statistical measurements on these 42 events.

\par
\begin{figure}[!t]
	\centering
	\includegraphics[width=0.45\textwidth]{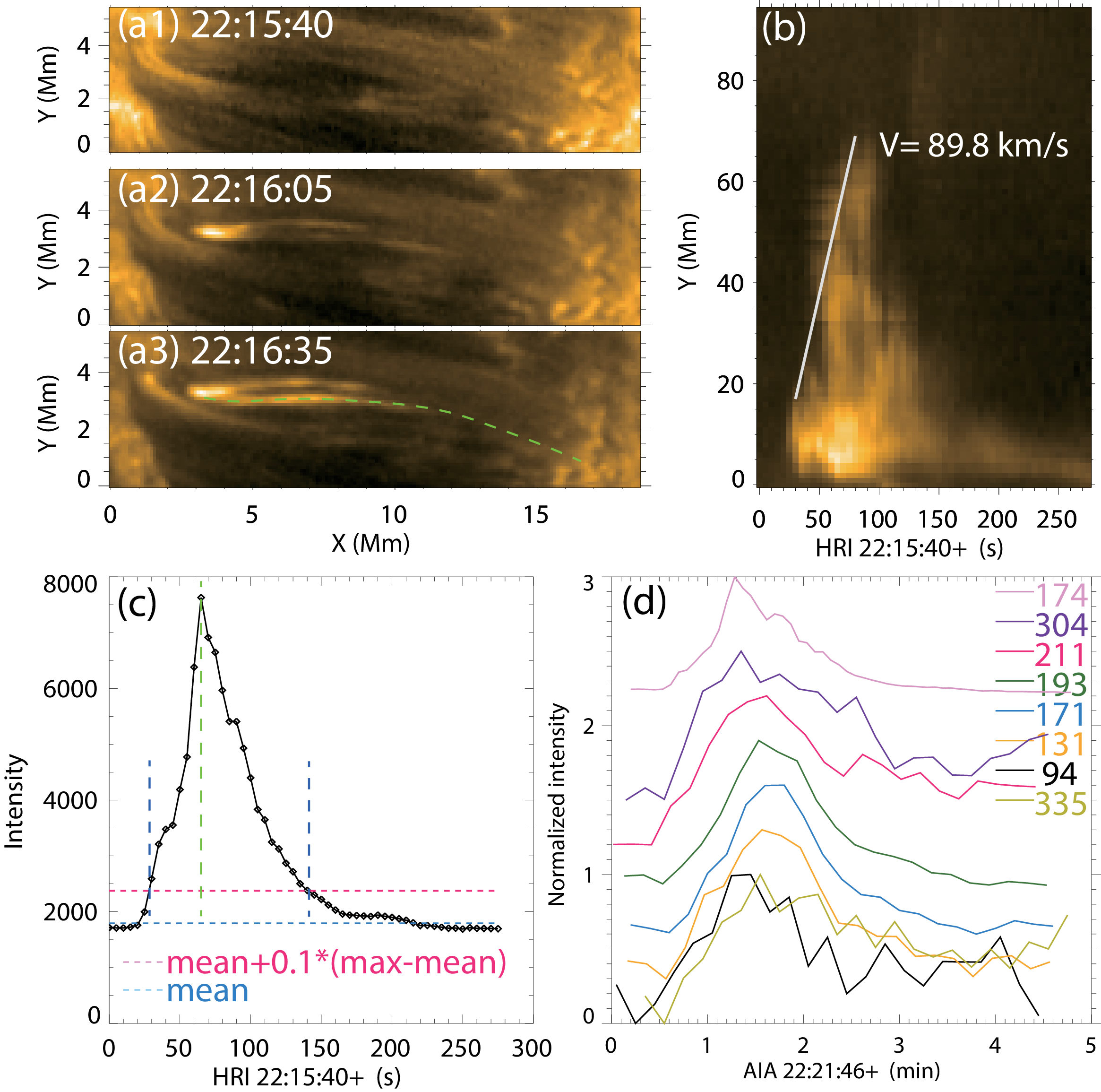}
	\caption{Case-analysis methodology.
		Panel (a) shows the evolution of Case-13, the green dashed line marks the trajectory of the brightening loop.
		Panel (b) presents the T--S map, constructed along the green dashed trajectory in panel (a3); the white solid line indicates the propagation velocity of the brightening front. The T--S map is used to derive the characteristic timescale, $L_b$, propagation velocity, and other physical properties of the event.
		Panel (c) shows the schematic illustration of the timescale determination based on the light curve obtained by integrating the intensity along the selected loop path. The blue dashed line denotes the mean intensity of the lowest 50\% of pixels, and the red dashed lines indicate the reference baseline used to identify the brightening and intensity decreasing phases.
		Panel (d) displays the light curves from multiple AIA passbands and HRI$_{\mathrm{EUV}}$ passband.}
	\label{fig2}
\end{figure}

For each event, we define a sampling path along the loop and obtain the intensity distribution along the loop for each frame. This allows us to construct a two-dimensional T--S map that characterizes the evolution of a brightening along the loop with time (Fig.~\ref{fig2}). The T--S maps are used for the following purposes: extract light curves and measure timescales; determine the apparent propagation velocity of the brightening front along the loop (Fig.~\ref{fig2}(b)), and analyze the initial brightening location.
The HRI$_{\mathrm{EUV}}$ data used here have cadences of 3~s and 5~s. Out of 42 events,  37 were observed at 5~s cadence and 5 events at 3~s cadence (see Table~\ref{table1}). For the timescale measurements, we extracted light curves from a temporal window extending approximately one minute before and after each event (about 10 frames for the 5~s cadence data and about 20 frames for the 3~s cadence data), so that the event was located near the center of the analysed time interval. The light curves were obtained by integrating the intensity along the selected loop path. For a few events, however, the temporal window was  slightly adjusted because neighboring brightenings affected the pre-event or post-event background. As a result, the number of frames selected before and after the event was not always identical. This interval includes a representative background level for subsequent quantification. We define the event background intensity ``mean" as the average of the lowest 50\% of intensity samples within this interval. The start time is defined as the first time the intensity exceeds mean $+$ 0.1*(max$-$mean), where ``max'' is the maximum intensity in the interval.
The brightening phase ends when the intensity reaches its peak, after which the event enters the intensity decreasing phase. The intensity decreasing phase ends when the intensity decreases back to mean $+$ 0.1*(max$-$mean) (Fig.~\ref{fig2}(c)).
Figure~\ref{fig2}(d) shows the response curves of the HRI and the AIA passbands. 
The AIA passbands generally exhibit highly similar temporal evolution, with nearly co-temporal peaks. The vast majority of the 42 events exhibit this behavior, providing a critical observational constraint for determining loop-temperature characteristics (Section~\ref{sec:dis}).

\begin{figure*}[!t]
	\centering
	\includegraphics[width=0.8\textwidth]{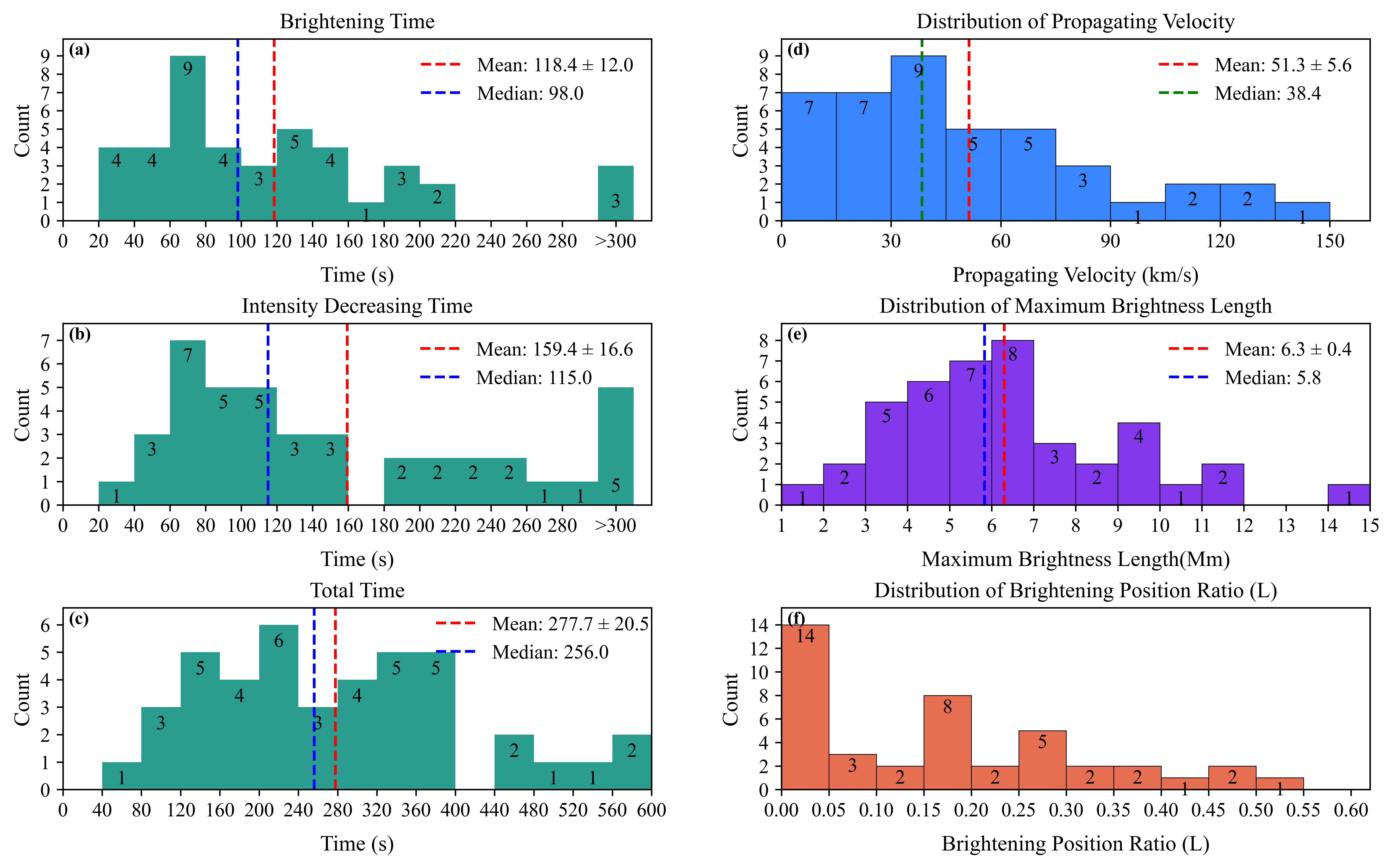}
	\caption{Statistical properties of 42 brightening events.
		Panels (a)--(c) show the distributions of the brightening time, intensity decreasing time, and total duration, the red dashed lines mark the mean values, and blue dashed lines mark the medians.
		Panel (d) shows the distribution of the propagation velocities of the brightening fronts. The red and green dashed lines denote the mean and median.
		Panel (e) displays the distribution of the maximum $L_b$, defined as the length of the contiguous brightened segment along the loop at the time of peak brightening. The red and blue dashed lines indicate the mean and median.
		Panel (f) presents the distribution of the initial brightening location, defined as the distance between the first clearly visible bright structure and the footpoint, normalized by the loop length. The detailed parameters of each event are listed in Table~\ref{table1}.
	}
	\label{fig3}
\end{figure*}

\par
Figures~\ref{fig3}(a)--(c) summarize the timescale statistics of the 42 events derived from the above definitions. The results indicate an impulsive character: brightening and intensity decreasing times show similar distributions, with most events falling within 20--160~s, and the highest occurrence is in the 60--80~s bin. A small number of events exhibit brightening and intensity decreasing times exceeding 300~s.
The mean brightening time is $118.4 \pm 12.0~\mathrm{s}$, and the mean intensity decreasing time is $159.4 \pm 16.6~\mathrm{s}$ (i.e., the latter is 1.35 times larger). The total duration is primarily distributed between 40 and 400~s, with a mean of $277.7 \pm 20.5~\mathrm{s}$. 
We note that the measured brightening time may exceed the true heating timescale. After energy input ceases, a finite time is required for energy and mass transport to reach the maximum observable brightening extent. Consequently, the true cooling-to-heating time ratio may exceed 1.35.

\par
Figure~\ref{fig3}(d) presents the apparent propagation velocities along the loop derived from the T--S maps. For each event, we tracked the main boundary of the brightening front in the T--S map and defined the propagation velocity as $v$ = $\Delta s$/$\Delta t$, where $\Delta s$ is the propagation along the loop over a time interval $\Delta t$.
The majority of events exhibit subsonic velocities in the range of 0--90~\kms, with the 0--15~\kms\ bin being the most populated. The mean velocity is $51.3 \pm 5.6$~\kms, and the median is 38.4~\kms. There are 7 events that yield $v$ $\approx$ 0, indicating that the brightenings remain stationary at their initial sites without significant propagation along the loop.

\par
Figure~\ref{fig3}(e) shows the distribution of $L_b$. To quantify the spatial ``impact range" of a single event, we define the maximum $L_b$ as the length of the contiguous enhanced intensity segment along the loop at the time of the strongest brightening.
In practice, we applied a threshold to the along-loop intensity profile for the frame in which the bright segment is most prominent. 
Pixels exceeding the threshold were classified as brightening, and the length of the contiguous above-threshold segment was converted to a physical length.
For a subset of events, strong loop superposition or an exceptionally bright background near the footpoints prevents robust application of this threshold. In these cases, we adopted a different method: we selected a pre-event frame where the loop is uniformly faint, subtracted it from the brightest frame to obtain an intensity enhancement map, and estimated the brightened length from this map.
The results show that most events have $L_b$ in the range of 3--11~Mm, with the highest frequency in the 5--8~Mm bin. The mean $L_b$ is $6.3 \pm 0.4$~Mm, and the median is 5.8~Mm. Given the viewing geometry and projection effects, these lengths likely represent lower limits to the true values. For a few events, the brightening segment reached the footpoints (Nos.~17, 25, and 33). These events have brightening ratios exceeding 0.9 (see Table.~\ref{table1}), meaning that the entire loop became bright. In these cases, the brightening length may have been limited by the loop length. We further note that bright footpoint backgrounds (e.g., a moss region) and multi-loop superposition introduce systematic uncertainty in the identification of bright segment boundaries. This uncertainty should be treated as a non-negligible error source when interpreting the length distribution and derived correlations.

\par
Figure~\ref{fig3}(f) summarizes the statistics of initial brightening locations. We define the initial brightening point as the location where a clear bright structure is first observed, quantified by the ratio of the distance from the footpoint to the initial brightening site relative to the loop length (L). The results show a pronounced footpoint preference: events are concentrated within 0--0.25~L, and the occurrence decreases monotonically with height within 0--0.5~L (Fig.~\ref{fig3}(f)).
This implies that, at least observationally, energy deposition and/or the first detectable brightening preferentially occur at low heights near the footpoint. Based on representative examples (Figs.~\ref{fig5}\&\ref{fig6}), we further categorize the events into four morphological classes:
(I) 4 cases, a brightening starts near a footpoint with no obvious loop-loop interaction, one example is shown in Figs.~\ref{fig5}(a1)--(a8);
(II) 16 cases, a brightening is footpoint triggered but shows migration of bright kernels adjacent to the footpoint or brightening of small loops (Figs.~\ref{fig5}(b1)--(b8));
(III) 20 cases, a brightening occurs along the loop, often accompanied by vigorous evolution of multiple intersecting or entangled small-scale loops (e.g., loop approach, loop separation, and along-loop propagation of bright segments (Figs.~\ref{fig6}(a1)--(a8));
(IV) 2 cases, a brightening with bidirectional jets consistent with reconnection outflows (Figs.~\ref{fig6}(b1)--(b8)).

\begin{figure}
	\centering
	\includegraphics[width=0.4\textwidth]{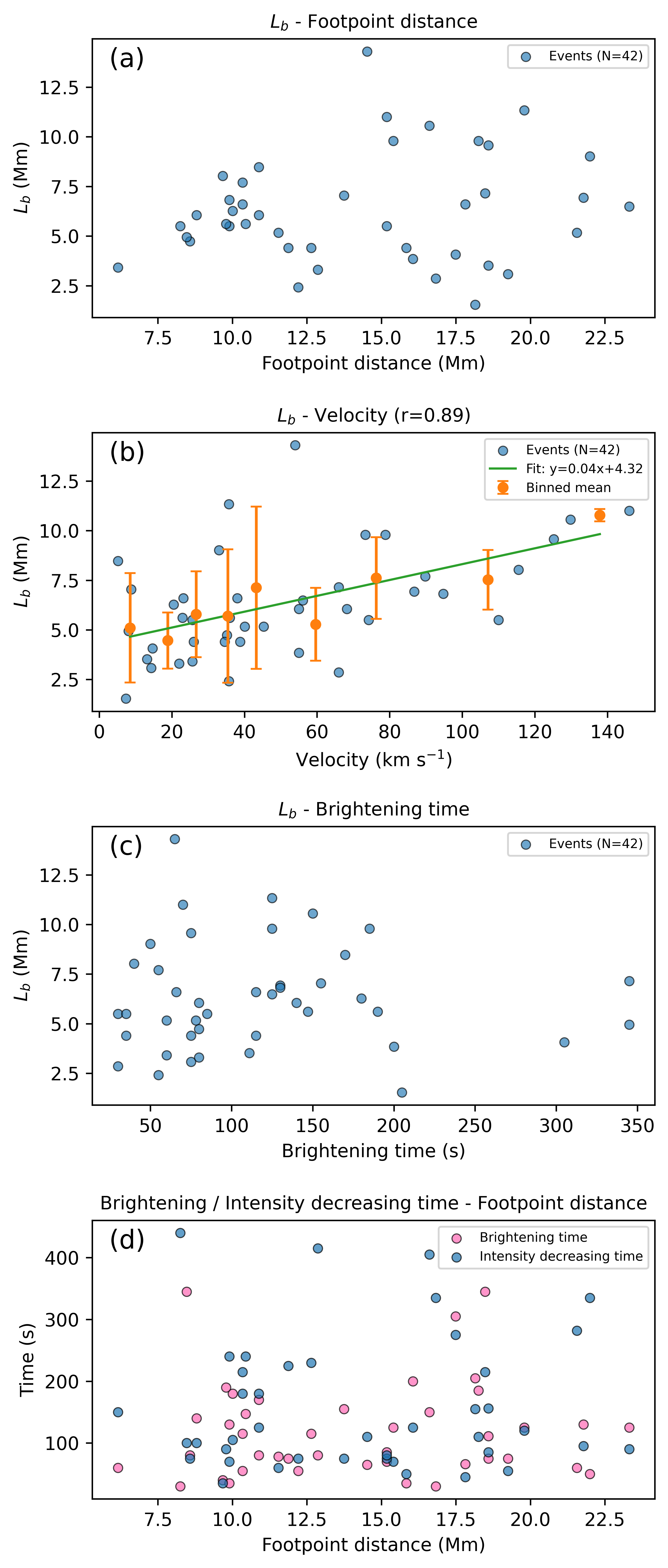}
	\caption{Two-dimensional correlation analysis.
		Panel (a) shows the correlation of $L_b$ and the footpoint distance.
		Panel (b) displays the correlation of $L_b$ and the propagation velocity. The green solid line is obtained from grouped mean data by linear fitting.
		Panel (c) presents the correlation of $L_b$ and the brightening time.
		Panel (d) presents the correlation of the brightening and intensity decreasing time and footpoint distance. The pink dots represent the brightening time, and the blue dots represent the intensity decreasing time.
	}
	\label{fig4}
\end{figure}

\begin{figure}[!t]
	\centering
	\includegraphics[width=0.5\textwidth]{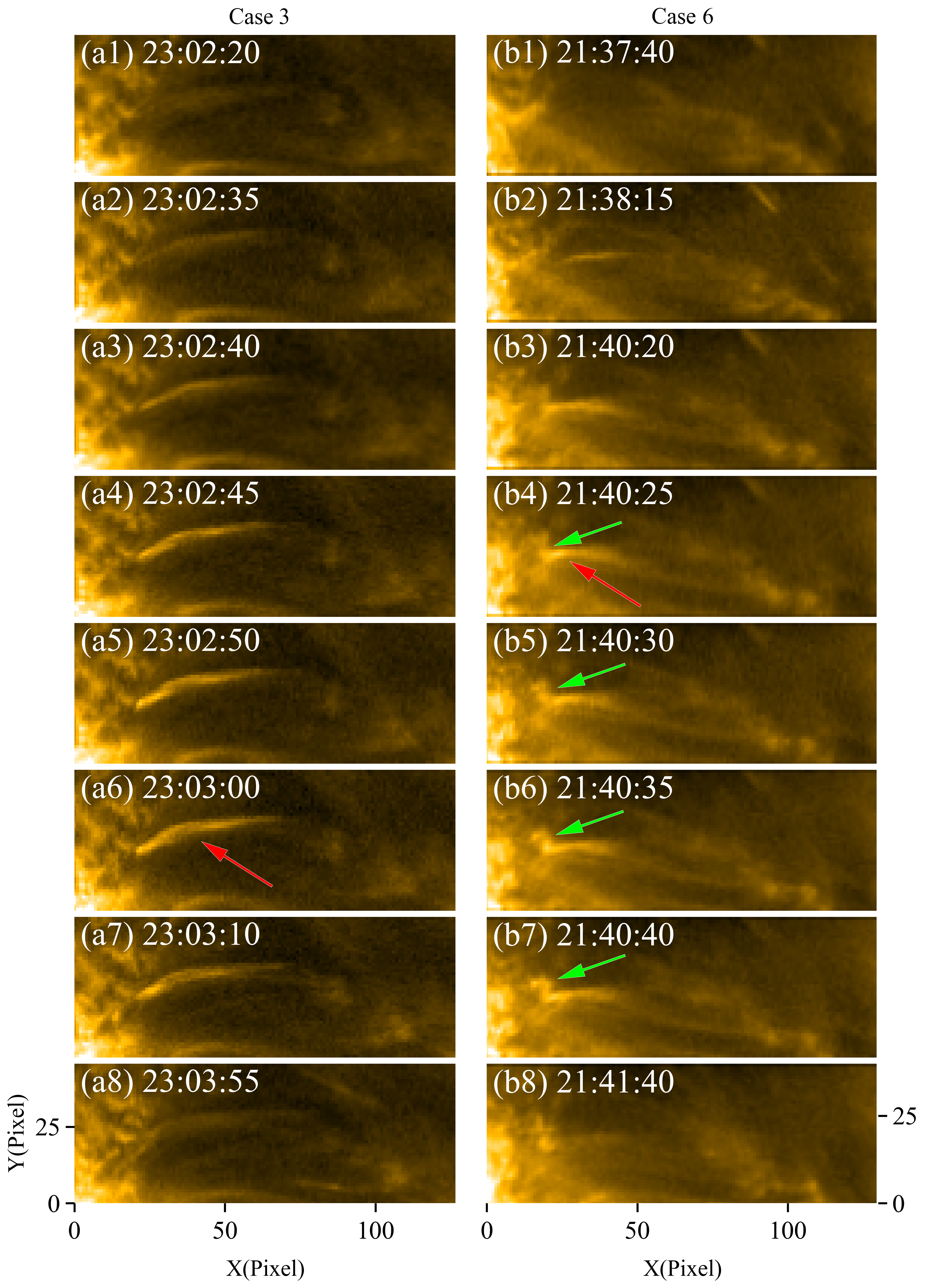}
	\caption{Evolution detail of brightening events.
		Panels (a1)--(a8) present the brightening process of Case-3; the event brightens at a footpoint, and the brightening front propagates along the loop over a finite distance.
		Panels (b1)--(b8) show the brightening process of Case-6; the event also brightens from the footpoint. As indicated by the green arrows, small, moving, brightening points are observed near the footpoint, which may provide evidence of reconnection between shorter and longer loops.
	}
	\label{fig5}
\end{figure}

\begin{figure}[!t]
	\centering
	\includegraphics[width=0.5\textwidth]{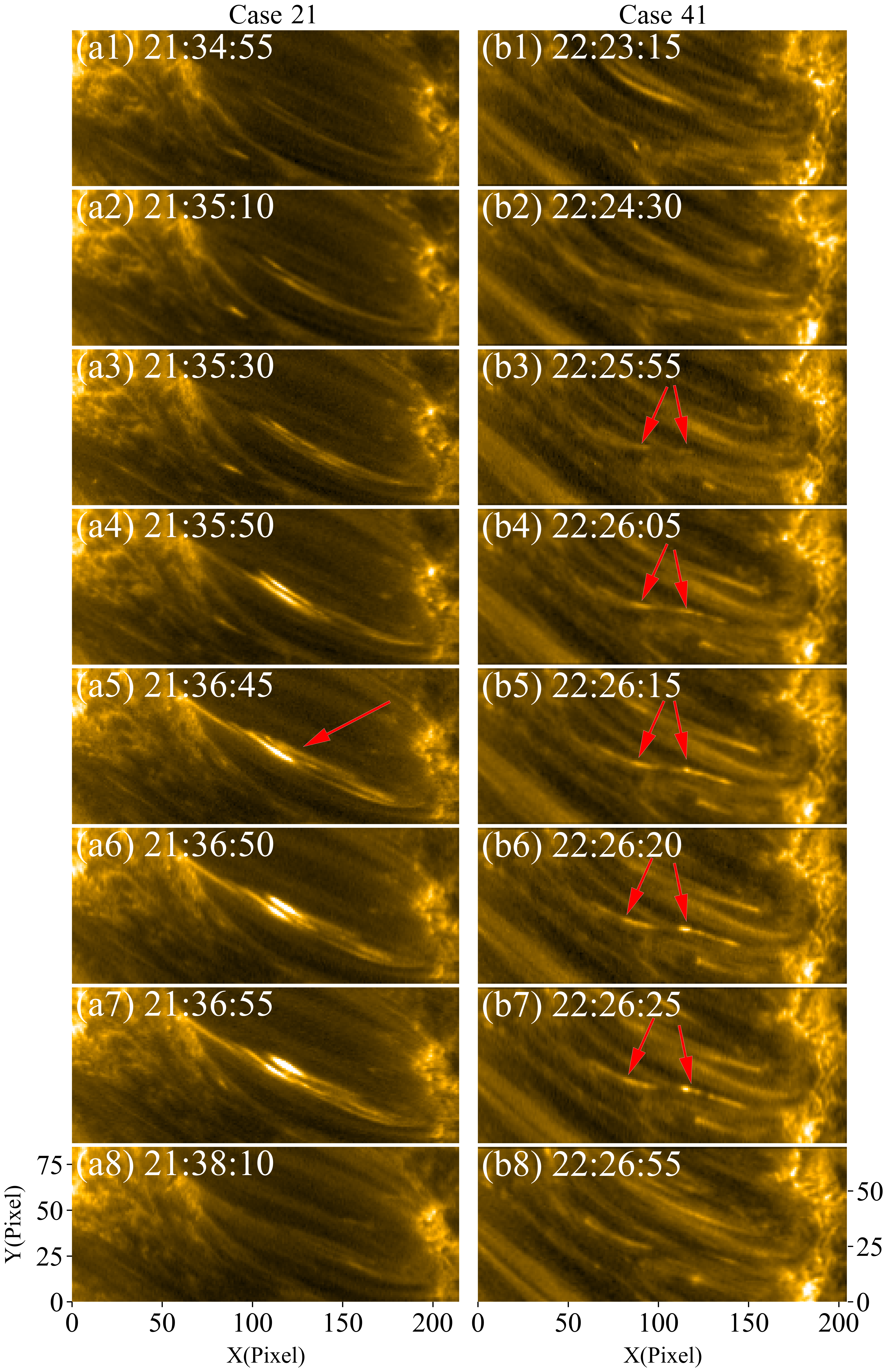}
	\caption{Evolution detail of brightening events.
		Panels (a1)--(a8) present the brightening process of Case-21, the event brightens near the loop apex, with multiple loops intersecting and separating during the evolution.
		Panels (b1)--(b8) show the brightening process of Case-41, as indicated by the red arrows, bidirectional jet-like features are observed during the brightening process, which may represent key observational signatures of magnetic reconnection between loops.
	}
	\label{fig6}
\end{figure}

\par
Based on the statistical results shown in Fig.~\ref{fig3}, we further analyze the correlations among these physical properties and examine their two-dimensional relationships in parameter space (Fig.~\ref{fig4}). To reduce the influence of scatter and emphasize overall behavior, we sort the independent variable and group the sample into bins of five events (Fig.~\ref{fig4}(b)). For each bin, we calculate the mean and standard deviation and plot error bars.
For pairs showing an apparent linear trend, we fit a linear regression to the binned means and report the corresponding correlation coefficient. The key results include:
In panel~(a), the scatter between $L_b$ and footpoint distance is large with no clear systematic trend, indicating a weak overall correlation. This suggests the spatial extent of a single brightening does not scale linearly with the loop’s geometric size.
In panel~(b), a clear positive trend exists between $L_b$ and propagation velocity, i.e., faster events tend to produce longer brightening segments.
In panel~(c), $L_b$ shows no significant correlation with heating time, implying that the duration of energy release and the resulting observable $L_b$ can be statistically decoupled.
In panel~(d), neither the brightening time nor the intensity decreasing time shows a clear correlation with the footpoint distance.

\begin{figure*}[ht!]
	\centering
	\includegraphics[width=\textwidth]{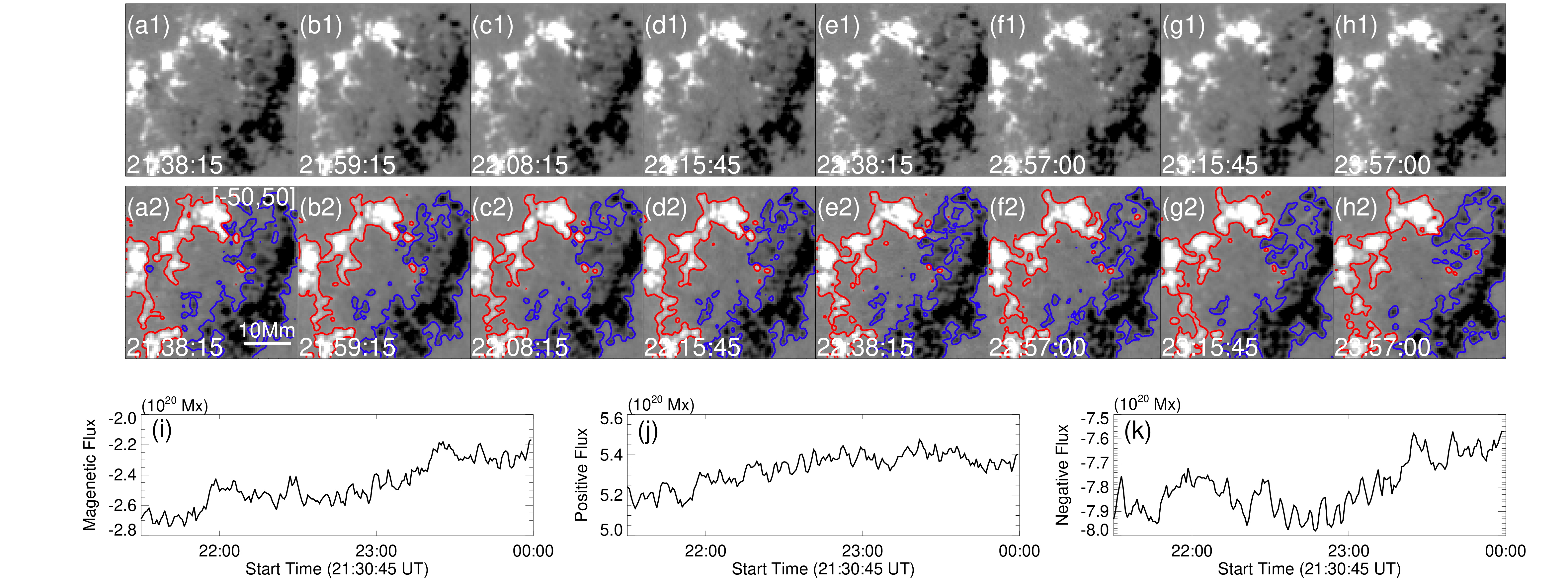}
	\caption{Magnetic field evolution.
		Panels (a1)--(h1) present the HMI magnetograms of the main region, where the brightening events occur.
		Panels (a2)--(h2), same as panel (a), but overlaid with $\pm$50~G magnetic field contours, facilitating the identification of magnetic polarity motion, emergence, and cancellation. Red and blue contours represent positive and negative magnetic polarities, respectively.
		Panels (i)--(k) show the temporal evolution of the total magnetic flux, positive magnetic flux, and negative magnetic flux of the region in panel (a1).
	}
	\label{fig7}
\end{figure*}

\section{Discussion}
\label{sec:dis}
The timescale statistics quantify the above thermal-response constraints into limits that can be directly used for physical inference. The characteristic times of the brightening and intensity decreasing phases are generally of the order of 100~s ($<300$~s), with the intensity decreasing phase systematically longer than the brightening phase (mean brightening time $t_{\mathrm{b}} = 118.4 \pm 12.0$~s, mean intensity decreasing time $t_{\mathrm{d}} = 159.4 \pm 16.6$~s, and $t_{\mathrm{d}}/t_{\mathrm{b}}\approx1.35$). Together with the highly synchronous responses across the AIA passbands, this ``short impulse plus slightly trailing decay'' pattern suggests a rapid-cooling scenario under transition-region or low-coronal conditions.	
To assess the relevant cooling mechanisms, we compare the observed timescales with conductive and radiative cooling estimates. The conductive cooling times can be written as\,\citep{2014LRSP...11....4R}
\begin{equation}
	\tau_c = \frac{3 n k_B T L^2}{\frac{2}{7}\kappa \, T^{7/2}},
\end{equation}
where $n$ is the particle density at the end of the heat pulse, $k_{\rm B}$ is the Boltzmann constant, $T$ is the plasma temperature, $L$ is the loop length, and $\kappa$ is the thermal conductivity. For representative parameters of our events, we adopt $L=15$~Mm, $n=10^{10}~{\rm cm^{-3}}$, $\kappa = 9 \times 10^{-7}~\mathrm{erg\,s^{-1}\,cm^{-1}\,K^{-7/2}}$, and $T=0.4$~MK, which gives $\tau_c \approx 3.5\times10^5$~s. This value is far longer than the observed intensity decreasing timescale (with an order of 100~s), indicating that classical conductive cooling is unlikely to be the dominant mechanism controlling the rapid intensity decreasing of these events. By contrast, the radiative cooling times can be written as\,\citep{1997ApJ...478..799C,2004ApJ...605..911C,2013ApJ...771...21W,2014LRSP...11....4R}
\begin{equation}
	\tau_R=\frac{3k_{\rm B}T}{n\Lambda(T)},
\end{equation}
where $n$ is the plasma density and $\Lambda(T)$ is the optically thin radiative loss function. For $T=0.4$~MK, $n=10^{10}~{\rm cm^{-3}}$, and $\Lambda(T)=8.87\times10^{-17}T^{-1}$\,\citep{2008ApJ...682.1351K}, we obtain $\tau_R\approx80$~s, which is close to the observed intensity decreasing timescale. The rapid intensity decreasing is therefore more consistent with radiative cooling than with classical conductive cooling.
In a relatively high-density environment, the redistribution of energy and mass along the magnetic field, such as enthalpy-flux regulation, may still act together with radiation to shape the decay phase, yielding a statistical preference for $t_{\mathrm{d}}\gtrsim t_{\mathrm{b}}$. More importantly, these timescales provide an order-of-magnitude constraint on the ``visibility time interval'': the effective duration $\tau$ over which the brightened structure maintains an identifiable contrast above the background is unlikely to be much larger than 100~s; otherwise, a substantially longer duration distribution would be expected. We therefore take $\tau \approx 100$~s as a key premise for subsequent interpretations of spatial extension and correlations.

\par
The velocity statistics provide dynamical constraints on ``field-aligned responses or transport'', but their physical meaning should be interpreted with caution. The apparent propagation velocities of the intensity front are mainly distributed in the range of 0--90~\kms\ (mean = $51.3 \pm 5.6$~\kms, median = 38.4~\kms). For a characteristic plasma temperature of 0.4~MK, the adiabatic sound speed is
\begin{equation}
	c_s=\sqrt{\frac{\gamma k_{\rm B} T}{\mu m_{\rm p}}},
\end{equation}
where $\gamma$ is the adiabatic index, $k_{\rm B}$ is the Boltzmann constant, $T$ is the plasma temperature, $\mu$ is the mean molecular weight, and $m_{\rm p}$ is the proton mass\,\citep{2019A&A...628A.133K}. Assuming a fully ionized plasma, we adopt $\gamma=5/3$ and $\mu=0.6$, yielding $c_s \approx 100$~\kms.
Most of the measured apparent propagation velocities are therefore below the local sound speed. Because these velocities are inferred from the apparent displacement of the intensity front in T--S maps, they may trace the front of a thermal disturbance, a pressure-gradient-driven field-aligned response (including density enhancements regulated by enthalpy flux that can brighten multiple passbands nearly simultaneously), or an apparent propagation produced by sequential energization of multiple unresolved strands. The coexistence of subsonic velocities and $v\approx0$ events favors a relatively gentle field-aligned response or transport, rather than strong shock-like high-velocity propagation as a ubiquitous scenario; meanwhile, the near-zero-velocity events suggest that not all brightenings require substantial field-aligned extension, and that localized energy deposition followed by predominantly local radiation and decay can also produce the observed evolution.

\par
Against the backdrop of an effective time interval of $\tau\approx100~\mathrm{s}$ and subsonic field-aligned responses, the two-parameter relationships provide key evidence linking timescales, velocities, and spatial extension. We find that the visible brightened segment in a single event is typically only a few Mm in length (concentrated in 3--11~Mm, with mean = $6.3 \pm 0.4$~Mm). The overall correlations of $L_b$ with the geometric scale of the loop (footpoint distance) and with the brightening timescale are weak, whereas a clearer positive trend is found with the propagation velocity ($v$). This suggests that the energy deposition relevant to the observable brightening is confined to a relatively limited spatial extent and is not strongly tied to the loop length, and that the transport of the brightening front plays a dominant role in setting $L_b$. These properties provide a plausible criterion for interpreting the energy transport in terms of enthalpy flux, i.e., energy carried along the field lines by enthalpy flux; given the high-density transition-region environment, radiative cooling is then expected to be important as well. Such apparent propagation may also reflect motion of the heating location; for example, \citet{2025ApJ...985...17Z} reported an observational case in which the unbraiding sites (heating location) migrated along the braiding axis at $\sim$45~\kms, consistent with a heating scenario involving multiple interacting loops.

\par
The onset locations of the brightenings show a statistically significant preference for the footpoints, with a peak at 0--0.25~L and a monotonic decrease with height. Based on the fine-scale properties at the heating site, the events can be grouped into four categories: (I) 4 events, starts at the footpoints without obvious inter-loop interactions; (II) 16 events, footpoint-triggered and exhibits migrating brightenings, which may reflect reconnection between a long loop and shorter loops rooted in the legs, where reconnection heats the plasma (producing the observed brightening) and the post-reconnection structures subsequently move under magnetic tension, leading to the apparent migration; (III) 20 events, brightens along the loop and shows braided, multi-loop evolution, in which crossing structures may undergo a sequence of reconnection episodes, and the moving post-reconnection loops can again satisfy reconnection-favorable geometries with other loops\,\citep{mondal2025relationshipnanoflareenergydelay}; (IV) 2 events, shows along-loop brightenings accompanied by bidirectional jets, which constitutes one of the key observational signatures of magnetic reconnection. These results establish an observationally grounded ``heating location'' constraint: for our sample, the earliest visible energy deposition or radiative enhancement preferentially occurs near the footpoints, rather than at the apex or at random locations along the loop. We emphasize that this ``footpoint preference'' is not driven by a few case studies but holds for the full sample under a uniform event definition and measurement procedure, and thus should be regarded as a basic constraint that viable mechanisms must satisfy. This is broadly consistent with the conclusions of\,\citet{2001ApJ...550.1036A}, who showed that extending uniformly heated models into the transition region can lead to unrealistic results (e.g., an excessively thick transition-region layer and column depths far exceeding observations), whereas footpoint-heating models with a scale height of $\sim$13~Mm can better reproduce the observed temperature distribution and flux profiles. More generally, coronal loop heating models spanning footpoint heating, apex heating, multi-site heating, and quasi-uniform heating can each reproduce certain observational signatures\,\citep{2001ApJ...550.1036A,2002ApJ...580..566R,2020A&A...644A.130C}, suggesting that the heating location in coronal loops may not exhibit a universal preference.

\par
Compared with typical coronal loops at $T>1~\mathrm{MK}$, the TR loop brightenings studied here evolve on shorter  timescales. In these loops, both brightening and intensity decreasing generally occur within about 100~s ($<300~\mathrm{s}$). This difference should not be attributed to loop length alone. In general, cooling timescales can depend on loop length when conductive cooling is important. For the representative parameters of our events, however, the classical conductive cooling timescale is much longer than the observed intensity decreasing timescale, whereas the radiative cooling timescale is much closer to the observations. This suggests that thermal conduction is unlikely to dominate the cooling of these TR loop brightenings, and that their rapid intensity decreasing is more likely governed primarily by radiative cooling under relatively cool and dense plasma conditions. 
In this sense, the cooling evolution in our sample is expected to be less sensitive to loop length. For hotter coronal loops, thermal conduction is often expected to play an important role during the early stage of the cooling evolution after the heating phase ends\,\citep{2014LRSP...11....4R}. Thus, the early cooling behavior of such coronal loops is generally more closely related to loop length than that of the events studied here.	
\par
By contrast, coronal loop studies commonly find heating phases lasting from minutes to tens of minutes, followed by cooling over tens of minutes to hours and producing systematic, temperature-dependent delays across multiple passbands\,\citep{2014LRSP...11....4R,2015A&A...583A.109L}. The review by \citet{2014LRSP...11....4R} further noted that hot active-region coronal loops are likely highly multistranded and may be continuously energized by heating events (nanoflares), rather than undergoing a single heating episode followed by free cooling. Such repeated small-scale energy injections can maintain high temperatures and emission measures on average; even if individual strand cools rapidly, the out-of-phase superposition can yield a long-duration radiative evolution at macroscopic scales\,\citep{2014ApJ...783...12U,2024NatAs...8..706L}. 
In addition, the temperature dependence of the radiative loss function should also be taken into account. Around and below about 0.5~MK, the radiative loss function rises rapidly\,\citep{2008ApJ...682.1351K}, which favors faster radiative cooling, whereas at higher temperatures its variation becomes shallower, leading to a longer radiative cooling timescale. Therefore, the longer cooling timescales commonly observed in hotter coronal loops likely reflect not only multistranded or repeatedly heated conditions, but also the intrinsic temperature dependence of the radiative loss function. A multistranded interpretation cannot be excluded for the TR loops studied here. In our statistical results, the apparent brightening sites are often concentrated near the footpoints (Fig.~\ref{fig3}(f)). If an observed loop actually consists of multiple unresolved strands, then these strands would need to be located very close to one another, with intersections or heating sites clustered within a small spatial range and triggered nearly simultaneously, so that the resulting emission still appears as a single impulsive brightening. Under the current instrumental resolution, such a scenario would observationally be expected to show a coherent loop-like morphology, a compact, localized brightening onset, and little clearly separable transverse substructure. Higher-resolution observations will be needed to determine whether such fine substructure is present in this kind of transition-region loops.

\par
Consistent with these timescale differences, intensity disturbances or thermal front propagation in coronal loops are often tens to $\sim$100+~\kms, comparable to slow-mode wave velocities or thermal-front velocities in $\sim$1~MK plasma\,\citep{2014LRSP...11....4R}, whereas the apparent propagation in TR loop brightenings more frequently lies in a subsonic and comparatively low-velocity regime. Overall, this contrast, TR loops: seconds to minutes and low velocities vs. coronal loops: minutes to hours and higher velocities, suggests different heating and transport regimes. This difference is less likely to be mainly controlled by loop length, as neither the brightening time nor the intensity decreasing time shows a clear correlation with the footpoint distance (Fig.~\ref{fig4}(d)). In this sense, TR loops may be more consistent with a single or brief impulsive response, whereas coronal loops more likely reflect multiple heating impulses and prolonged cooling. At the same time, however, we cannot exclude the possibility that some TR loop brightenings involve multiple impulsive heating episodes in multiple strands, as discussed above.

\par
We interpret the energy and mass-transport mechanism in these brightenings in terms of enthalpy flux, i.e., a field-aligned mass flow with finite velocity that transports energy along the loop as the plasma cools. The enthalpy flux mechanism provides an alternative diagnostic approach for transition-region plasma. In this framework, the energy transported by a field-aligned mass flow is assumed to be radiated away locally, leading to an energy balance condition: $F_h \approx n^2 \Lambda(T) l$. The enthalpy flux is given by
$F_h = \frac{5}{2} p v$,
with $p = \eta n k_B T$ and $\eta \approx 2$, which yields
$F_h = 5 n k_B T v$. Rearranging the energy balance yields a diagnostic expression:
\begin{equation}
T \approx \frac{n\, l\, \Lambda(T)}{5\, k_B\, v},
\end{equation}
where $n$ is the density, $l$ is the length of the brightening segment, $v$ is the flow velocity, and $\Lambda(T)$ is the radiative loss function\,\citep{1981ApJ...247..686R,2008ApJ...682.1351K}. 

If we adopt $\Lambda(T) = 8.87 \times 10^{-17}\ T^{-1}$ for $10^{4.97} < T \le 10^{5.67}~\mathrm{K}$\,\citep{2008ApJ...682.1351K}, take $k_B$ as the Boltzmann constant, assume a typical transition-region or low-coronal density $n = 10^{10}~\mathrm{cm^{-3}}$, use the median brightening length $l_m$ = 6.3~Mm, and the median propagation velocity $v_m$ = 38.4~\kms, we obtain $T \approx$ 0.46~MK. This temperature estimate is consistent with the typical transition-region temperature range and well-matched to the response profiles of the AIA passbands (Fig.~\ref{fig2}(d)). Furthermore, if the temperature is known, this method can be readily extended to perform plasma density diagnostics. We note that the results presented here are only rough estimates and, thus, should be interpreted with caution. This is because a precise calculation for any particular loop cannot be achieved with the current dataset, owing to substantial diversity in plasma density and the radiation loss function. That said, this method proposes a novel approach to temperature and density diagnostics in the transition region and low corona, which merits further investigation and in-depth validation in future studies.

\par
The HMI magnetic field data provide evidence for field evolution consistent with footpoint-driven intermittent energy release, although the limitations in temporal and spatial resolution prevent a strict one-to-one correspondence between individual events and specific magnetic features. Figures~\ref{fig7}(a1)--(h1) show magnetograms of the primary source region over 21:30:45--24:00:00~UT (about 2.5~hrs), and Figs.~\ref{fig7}(a2)--(h2) further overlay $\pm50~\mathrm{G}$ contours to highlight the evolution of the magnetic structure. The contour morphology changes markedly over time and is accompanied by patch motions, coalescence, and fragmentation, as well as local emergence and cancellation, indicating ongoing flux redistribution and topological restructuring during the observing time. Correspondingly, Figs.~\ref{fig7}(i)--(k) show the flux evolution within the region in Fig.~\ref{fig7}(a1): the total flux changes from $-2.8\times10^{20}~\mathrm{Mx}$ to $-2.2\times10^{20}~\mathrm{Mx}$; the positive flux increases from $5.1\times10^{20}~\mathrm{Mx}$ to $5.5\times10^{20}~\mathrm{Mx}$; and the negative flux evolves from $-8.0\times10^{20}~\mathrm{Mx}$ to $-7.6\times10^{20}~\mathrm{Mx}$. Notably, multiple short-lived peaks are superposed on the overall trends, implying that the flux evolution is not smoothly monotonic but rather composed of intermittent small-scale variations that phenomenologically echo the impulsive nature of the brightenings. The evolving contours and the multi-peaked flux curves suggest that processes such as flux cancellation and magnetic braiding may have operated during this interval, providing necessary conditions for magnetic energy release. Nevertheless, we emphasize that this inference should be treated with caution, because the complexity of the field evolution, together with HMI's spatial resolution and projection effects, makes it difficult to establish a strict one-to-one association between a given brightening and a specific cancellation and coalescence episode.

\section{Summary} \label{sec:con}
Based on coordinated observations from EUI/SO and AIA/SDO, we present a statistical investigation of transient brightening events occurring in transition-region loops. Taking advantage of the dual-viewpoint observing geometry and a unified T--S analysis, we identify 42 events for statistical analysis. After correcting for instrumental time-lag, the light curves of all AIA EUV passbands exhibit highly simultaneous temporal evolution: the brightening and subsequent intensity decreasing occur nearly simultaneously, with well-aligned peak times across passbands. The observations are consistent with impulsive plasma responses in the transition-region to low-coronal temperature range, where the radiative enhancement is primarily modulated by density variations. Statistical analyses of the characteristic timescales further reinforce this interpretation. The events display clear impulsive signatures, with a mean brightening time of $118.4 \pm 12.0~\mathrm{s}$ and a mean intensity decreasing time of $159.4 \pm 16.6~\mathrm{s}$. Such temporal asymmetry favors rapid cooling influenced by radiative losses and field-aligned enthalpy flux regulation.

\par
The apparent propagation velocity of the intensity fronts predominantly lies within the subsonic range of 0--90~\kms, with an average value of $51.3 \pm 5.6$~\kms, while the visible $L_b$ in individual events is typically limited to only a few megameters (3--11~Mm). The $L_b$ shows only weak dependence on the geometric loop length, but exhibits a clearer correlation with the propagation velocity. This behavior is consistent with a scenario in which field-aligned transport within a limited effective time interval governs the spatial extent of the brightening, i.e., $L_b$ $\sim$ $v$$\tau$, where $\tau$ is constrained by a rapid cooling process on the order of 100~s. Based on the enthalpy-flux mechanism and key parameters (e.g., $v_m$ and $l_m$) of brightening events, we propose a potential diagnostic method for the temperature and density of the transition region. The temperature derived from this method is consistent with the typical temperature characteristics of the transition region and also matches the response profiles of AIA passbands. The initiation sites of the brightenings show a pronounced preference for loop footpoints, with most events originating within the 0--0.25~L. Time series of HMI/SDO magnetograms reveal significant magnetic flux evolution and the motion of small-scale magnetic elements in the source regions. These magnetic characteristics provide a plausible environmental context for magnetic flux cancellation, footpoint-driven magnetic braiding, and interchange reconnection as potential triggering mechanism(s). Although it is not yet possible to unambiguously associate individual brightening events with specific magnetic activities, the observed magnetic evolution and the statistical properties of the events are physically self-consistent.

\par
Taken together, our results suggest that transition-region loop brightenings are driven by short-lived, impulsive energy injections near the loop footpoints, followed by rapid field-aligned responses and cooling. The combined constraints from timescale, propagation velocity, $L_b$, and initiation location provide important quantitative benchmarks for future high-resolution observations and numerical modeling of impulsive energy release processes in the fine structures of the solar atmosphere.

\begin{acknowledgements}
We are grateful to the anonymous referee for the critical and constructive comments that helped improve the manuscript.
This work is supported by the National Natural Science Foundation of China (Nos. 42230203, 42174201) and the National Key R\&D Program of China (2021YFA0718600).
M.M. acknowledges DFG grant WI 3211/8-2, project number 452856778, and was supported by the Brain Pool program funded by the Ministry of Science and ICT through the National Research Foundation of Korea (RS-2024-00408396).
Z.H. and M.M. acknowledge the support by the International Space Science Institute (ISSI) in Bern, through ISSI International Team project \#24-605 (Small-scale magnetic flux ropes under the microscope with Parker Solar Probe and Solar Orbiter).
M.M. thanks the International Space Science Institute, Bern, Switzerland, for the Visiting Scientist grant.
Solar Orbiter is a space mission of international collaboration between ESA and NASA, operated by ESA. The EUI instrument was built by CSL, IAS, MPS, MSSL/UCL, PMOD/WRC, ROB, LCF/IO with funding from the Belgian Federal Science Policy Office (BELSPO/PRODEX PEA 4000134088); the Centre National d'\'Etudes Spatiales (CNES); the UK Space Agency (UKSA); the Bundesministerium f\"ur Wirtschaft und Energie (BMWi) through the Deutsches Zentrum f\"ur Luft- und Raumfahrt (DLR); and the Swiss Space Office (SSO).
The AIA and HMI data are used by courtesy of NASA/SDO, the AIA and HMI teams, and JSOC.
\end{acknowledgements}

\bibliography{references}
\bibliographystyle{aa}

\begin{appendix}	
\FloatBarrier
\onecolumn
\section{Detailed parameters}
\begin{table*}[ht!]
	\caption{\label{table1}Detailed parameters of 42 brightening events sorted by category.}
	\centering
	\begin{tabular}{ccccccccc}
		\hline
		Num & Category & \makecell{Velocity\\(\kms)} & \makecell{Brightening\\time (s)} & \makecell{Intensity decreasing\\time (s)} & \makecell{Duration (s)} & \makecell{Temporal\\cadence (s)} & \makecell{Footpoint\\distance (Mm)} & \makecell{Brightening\\ratio} \\
		\hline
		1  & I   & 8.8   & 155 & 75  & 230 & 5 & 13.75 & 0.53 \\
		2  & I   & 55.0  & 200 & 125 & 325 & 5 & 16.06 & 0.33 \\
		3  & I   & 25.7  & 35  & 70  & 105 & 5 & 9.90  & 0.68 \\
		4  & I   & 38.0  & 66  & 45  & 111 & 3 & 17.82 & 0.36 \\
		5  & II  & 35.2  & 80  & 75  & 155 & 5 & 8.58  & 0.63 \\
		6  & II  & 20.5  & 180 & 105 & 285 & 5 & 10.01 & 0.62 \\
		7  & II  & 5.2   & 170 & 180 & 350 & 5 & 10.89 & 0.65 \\
		8  & II  & 129.8 & 150 & 405 & 555 & 5 & 16.61 & 0.57 \\
		9  & II  & 14.7  & 305 & 275 & 580 & 5 & 17.49 & 0.32 \\
		10 & II  & 94.8  & 130 & 240 & 370 & 5 & 9.90  & 0.84 \\
		11 & II  & 14.4  & 75  & 55  & 130 & 5 & 19.25 & 0.20 \\
		12 & II  & 68.2  & 80  & 125 & 205 & 5 & 10.89 & 0.53 \\
		13 & II  & 89.8  & 55  & 180 & 235 & 5 & 10.34 & 0.72 \\
		14 & II  & 35.8  & 55  & 75  & 130 & 5 & 12.21 & 0.20 \\
		15 & II  & 34.6  & 75  & 225 & 300 & 5 & 11.88 & 0.41 \\
		16 & II  & 110.0 & 30  & 440 & 470 & 5 & 8.25  & 0.68 \\
		17 & II  & 115.5 & 40  & 35  & 75  & 5 & 9.68  & 0.94 \\
		18 & II  & 25.7  & 60  & 150 & 210 & 5 & 6.16  & 0.71 \\
		19 & II  & 36.0  & 147 & 240 & 387 & 3 & 10.45 & 0.77 \\
		20 & II  & 45.3  & 78  & 60  & 138 & 3 & 11.55 & 0.44 \\
		21 & III & 78.8  & 125 & 70  & 195 & 5 & 15.40 & 0.61 \\
		22 & III & 7.3   & 205 & 155 & 360 & 5 & 18.15 & 0.08 \\
		23 & III & 66.0  & 345 & 215 & 560 & 5 & 18.48 & 0.55 \\
		24 & III & 125.2 & 75  & 85  & 160 & 5 & 18.59 & 0.57 \\
		25 & III & 73.3  & 185 & 110 & 295 & 5 & 18.26 & 0.98 \\
		26 & III & 33.0  & 50  & 335 & 385 & 5 & 22.00 & 0.44 \\
		27 & III & 86.8  & 130 & 95  & 225 & 5 & 21.78 & 0.34 \\
		28 & III & 56.1  & 125 & 90  & 215 & 5 & 23.32 & 0.43 \\
		29 & III & 26.0  & 35  & 50  & 85  & 5 & 15.84 & 0.26 \\
		30 & III & 146.0 & 70  & 75  & 145 & 5 & 15.18 & 0.83 \\
		31 & III & 38.8  & 115 & 230 & 345 & 5 & 12.65 & 0.83 \\
		32 & III & 66.0  & 30  & 335 & 365 & 5 & 16.83 & 0.15 \\
		33 & III & 54.0  & 65  & 110 & 175 & 5 & 14.52 & 1 \\
		34 & III & 35.8  & 125 & 120 & 245 & 5 & 19.80 & 0.63 \\
		35 & III & 55.0  & 140 & 100 & 240 & 5 & 8.80  & 0.81 \\
		36 & III & 23.2  & 115 & 215 & 330 & 5 & 10.34 & 0.72 \\
		37 & III & 74.3  & 85  & 80  & 165 & 5 & 15.18 & 0.34 \\
		38 & III & 40.1  & 60  & 282 & 342 & 3 & 21.56 & 0.23 \\
		39 & III & 13.2  & 111 & 156 & 267 & 3 & 18.59 & 0.22 \\
		40 & III & 22.9  & 190 & 90  & 280 & 5 & 9.79  & 0.63 \\
		41 & IV  & 8.1   & 345 & 100 & 445 & 5 & 8.47  & 0.73 \\
		42 & IV  & 22.0  & 80  & 415 & 495 & 5 & 12.87 & 0.20 \\
		\hline
	\end{tabular}
	\tablefoot{The second column shows the classification of the detailed brightening behavior of the events: (I) a brightening starts near a footpoint with no obvious loop–loop interaction; (II) a brightening is footpoint triggered but exhibits migration of bright kernels adjacent to the footpoint or brightening of small loops; (III) a brightening occurs along the loop and is often accompanied by vigorous evolution of multiple intersecting or entangled small-scale loops; and (IV) a brightening with bidirectional jets consistent with reconnection outflows. The third to ninth columns list, respectively, the propagation velocity, the brightening time, the intensity decreasing time, the total duration, temporal cadence, the footpoint distance, and the brightening ratio for each event. The brightening ratio is the fraction of the loop length occupied by the brightened segment at its maximum extent.}
\end{table*}

\end{appendix}

\end{document}